\documentclass[DIV=12]{scrartcl}
\usepackage{amsmath,amssymb}
\usepackage[
colorlinks=true,
linkcolor=blue,
citecolor=blue,
urlcolor=blue,
linktocpage=true
]{hyperref}
\usepackage[capitalise]{cleveref}
\usepackage{slashed}
\usepackage{bm}
\usepackage{braket}
\usepackage{ascmac}
\usepackage{mathrsfs}
\usepackage{cite}
\usepackage{graphicx}
\usepackage{mathtools}
\usepackage{color}

\newcommand{\hodge}{{\star}}
\newcommand{\tr}{\operatorname{tr}}

\crefname{section}{Sec.}{Secs.}

\begin{document}

\begin{titlepage}
\setkomafont{pagenumber}{\normalfont\color{white}}
\setkomafont{pageheadfoot}{\normalfont\normalcolor}
\thispagestyle{myheadings}
\markright{\rightline{KEK-TH-2282}}
\centering


\vspace*{100pt}
{\usekomafont{disposition}\huge Vector mesons on the wall}

\renewcommand{\thefootnote}{\fnsymbol{footnote}}

\vspace{30pt}
{\Large Ryuichiro Kitano$^{1,2,}$\footnote{Electronic address: \href{mailto:ryuichiro.kitano@kek.jp}{ryuichiro.kitano@kek.jp}} and Ryutaro Matsudo$^{1,}$\footnote{Electronic address: \href{mailto:ryutaro.matsudo@kek.jp}{ryutaro.matsudo@kek.jp}}}

\renewcommand{\thefootnote}{\arabic{footnote}}
\setcounter{footnote}{0}

\vspace{10pt}

\textit{\large ${}^{1}$KEK Theory Center, Tsukuba 305-0801, Japan}

\textit{\large ${}^{2}$Graduate University for Advanced Studies (SOKENDAI), Tsukuba 305-0801, Japan}

\vspace{30pt}
\textbf{\large Abstract}
\vspace{0pt}

\begin{abstract}
A domain-wall configuration of the $\eta'$ meson bounded by a string
(called a pancake or a Hall droplet) is recently proposed to describe
the baryons with spin $N_c/2$.
In order to understand its baryon number as well as the flavor quantum
number, we argue that the vector mesons (the $\rho$ and $\omega$ mesons)
should play an essential role for the consistency of the whole picture.
We determine the effective theory of large-$N_c$ QCD with $N_f$
massless fermions by taking into account a mixed anomaly involving
the $\theta$-periodicity and the global symmetry. 
The anomaly matching requires the presence of a dynamical domain wall on which a $U(N_f)_{-N_c}$ Chern-Simons theory is supported.
%
We consider the boundary conditions that should be imposed on the edge of the domain wall, and conclude that there should be a boundary term that couples the $U(N_f)_{-N_c}$ gauge field to the vector mesons.
We discuss the impact on physics of the chiral phase transition and the relation to the ``duality'' of QCD.
\end{abstract}

\end{titlepage}

\setcounter{page}{1}

\tableofcontents

\section{Introduction}

The low energy effective theory of QCD in the limit of the massless
$N_f$-flavor quarks is a non-linear sigma model corresponding to the
breaking $SU(N_f)_L\times SU(N_f)_R\rightarrow SU(N_f)_V$.  The 't~Hooft
anomaly~\cite{Hooft1980} for $SU(N_f)_L\times SU(N_f)_R\times U(1)_V$ is
matched by the Wess-Zumino-Witten (WZW)
term~\cite{Wess1971,Witten1983a}.
%
%
In the large $N_c$ limit~\cite{Hooft1974} of massless QCD, the axial
symmetry $U(1)_A$ is restored and the $\eta'$ meson is regarded as the
Nambu-Goldstone (NG) boson associated with the spontaneous $U(1)_A$
breaking~\cite{Witten1979,Veneziano1979}. The $\eta'$ field is,
therefore, described as a $U(1)$-valued pseudo-scalar field.  The WZW
term can be extended to include $\eta'$.
In the large $N_c$ limit, the baryons should be described as solitons
since the mass and the couplings of order $N_c$ and $1/N_c$,
respectively, are naturally explained~\cite{Witten1979a,Witten1983}.
Indeed, the Skyrmion configurations~\cite{Skyrme1961} of the pions are identified as the
baryons by the coupling between the external baryon gauge field
and the topological charge in the WZW term.

Recently, the notion of the 't~Hooft anomalies and their matching has
been extended to the discrete
symmetries~\cite{Kapustin2014b,Kapustin2014} and higher form
symmetries~\cite{Gaiotto2015}.  One of the important results is given in
Ref.~\cite{Gaiotto2017}, where a mixed anomaly involving time reversal
and the 1-form center symmetry is discussed in the $SU(N_c)$ Yang-Mills theory
at $\theta=\pi$.  The anomaly implies that if the confinement persists
at $\theta = \pi$, there are two degenerate vacua corresponding to the
spontaneous breaking of time reversal (or CP), and the effective theory
on a domain wall separating two vacua is an $SU(N_c)_1$ Chern-Simons
(CS) theory\footnote{
There are other logical options but an $SU(N_c)_1$
is the most plausible one~\cite{Hsin2019}. See also~\cite{Kitano2017} for an alternative argument for the existence of the phase transition.
}.  One can also consider the
't~Hooft anomaly for symmetries in the parameter space of the
theory~\cite{Cordova2020, Cordova2020a}.  In QCD, there is a mixed
anomaly involving the $\theta$-periodicity and the global symmetry
\cite{Cordova2020a}.
The matching of such anomalies provides new restrictions on the
effective theory especially for topologically non-trivial objects.


In the effective theory of QCD including $\eta'$, there are
topologically non-trivial configurations of $\eta'$ in addition to the
Skyrmions.
%
One of them is the domain-wall configuration which connects two minima
of the periodic potential of $\eta'$, where the potential is generated
at the next-to-leading order of the $1/N_c$ expansion. Although this
object is stable within the effective theory, it should not be so in
full QCD since there is no corresponding conserved charges.
The object to destabilize the domain wall can be identified as a string
around which $\eta'$ winds. Komargodski recently pointed out that the
domain wall bounded by a string, which is referred to as a pancake or a
Hall droplet, can have stable excited states that can be regarded as a spin-$N_c/2$
baryon~\cite{Komargodski2018a}.  This proposal is based on an
expectation that the effective theory on the domain wall is a CS theory,
which implies that there must be chiral edge modes on the boundary of
the domain wall.  Excited states on the edge with a unit $U(1)$ charge
can be regarded as baryons if we identify the $U(1)$ symmetry of the
chiral edge modes as the baryon symmetry.  It has been proposed that
this baryon can be understood as a chiral bag in a $1+2$ dimensional
strip using the Cheshire Cat principle~\cite{Ma2019}.

The existence of a CS theory on the domain wall is an expectation from
the spontaneous CP breaking in the Yang-Mills theory at $\theta=\pi$.
%
If we insert a interface of $\theta$ changing from $0$ to $2\pi$ in the
Yang-Mills theory, there appears a domain wall with a CS theory on it at
the location of the interface.
When massless quarks are added, the $\theta$
dependence is eliminated by a shift of $\eta'$, and thus a domain wall
where $\eta'$ shifts from a minimum of the potential to another
corresponds to the $\theta$ interface in the Yang-Mills theory.  This means
that it is expected that the domain wall of $\eta'$ supports a CS theory
on it.  
However, this explanation is rather heuristic and there are also
proposals that such a CS theory is replaced by a topologically trivial gapped theory when light quarks are added to the Yang-Mills theory~\cite{Karasik2020,Karasik2020a}.

The pancake story provides a new insight into the role of the vector
mesons ($\rho$ and $\omega$) as the dual
gluons~\cite{Kan2019,Ma2020,Karasik2020,Karasik2020a} in the sense of
the Seiberg duality in $\mathcal N=1$ supersymmetric
QCD~\cite{Seiberg1995}.
The interpretation of the vector mesons as the Seiberg dual gauge boson
has been discussed in literature~\cite{Seiberg2001, Harada1999,
Komargodski2011, Kitano2011, Abel2012, Kitano2012, Kitano2013} where the
vector mesons are identified as the Higgsed magnetic gauge bosons.
Indeed, the effective theory of the vector mesons can be formulated
consistently as the gauge theory of a hidden local symmetry in the
non-linear sigma model~\cite{Bando1985}.
In this formulation, the WZW term can be extended to include the vector
mesons by adding a set of gauge invariant operators~\cite{Fujiwara1985}.
The Lagrangian of the vector mesons on the domain wall is determined up
to four parameters. It is certainly interesting to note that the
postulated CS theory on the wall is $U(N_f)_{-N_c}$ whereas the vector
mesons can also form a $U(N_f)$ CS Higgs theory on the wall. It is
natural to suspect that there is some relation between these two
theories.
Since one can formally interpret the $U(N_f)_{-N_c}$ CS theory on the
wall as that of the (level-rank) dual gluon, there seems to be a deep
connection between the hidden local symmetry and the Seiberg duality
through the theory on the domain wall.
The topics of the baryon number of the pancake objects, the topological quantum
field theory on the wall, new 't~Hooft anomalies, and the Seiberg
duality seem to be all related, and it is quite important now to look
for the consistent picture and determine the effective (or dual) theory.

%

In this paper, we determine the effective theory of pions and $\eta'$
including the background $\theta$ term as well as background gauge
fields of global symmetries, $SU(N_f)_L \times SU(N_f)_R \times
U(1)_V$. We include the domain wall as the dynamical degree of freedom,
that is necessary to reproduce the cusp in the $\eta'$ potential.
We show that some topological quantum field theory should live on
the domain wall by using the anomaly matching involving the periodicity
of $\theta$. This confirms the heuristic explanation of the presence of
the $U(N_f)_{-N_c}$ theory on the domain wall.
We find that the anomalous coupling between $\eta'$ and the background
fields implies that the global $U(1)$ symmetry of the CS theory is
actually identified with the baryon symmetry, and the exited states with
a unit $U(1)$ charge belong to the correct representation of the flavor
symmetry. In addition, we propose a specific form of the coupling
between the vector meson fields and the gauge field of the CS theory on
the edge of the pancake in order to recover the gauge invariance of the
$SU(N_f)_L\times SU(N_f)_R$ gauge field, which must be maintained for
consistency.
From this effective theory, one can understand the real meaning of the
duality between the gluon and the vector mesons.
We discuss the implication of the behavior of the vector mesons near the
chiral phase transition.
The effective theory we derived is quite different from the proposals in \cite{Karasik2020,Karasik2020a}, where the theory on the pancake is claimed to be a CS Higgs theory of the vector mesons.
We argue that the CS theory, not CS Higgs theory, is required by the matching of the anomaly while the vector mesons mix with the gauge field of the CS theory on the edge of the pancake.
There are important differences in the formula of the baryon and flavor currents.
The existence of the CS theory on the wall will also be quite important in the discussion of the physics of the vector mesons such as the anomalous coupling among hadrons as well as how they approach to the chiral symmetric phase.

This paper is organized as follows: 
In \cref{sec:eta}, we determine the effective theory for $\eta'$
including the anomalous coupling to the background fields.  With an
appropriate counter term, the constant $\theta$ dependence is
compensated by a shift of $\eta'$.  However, we find that naive coupling
between $\eta'$ and the background fields satisfying this feature is not
adequate when we consider a violation of the $\theta$-periodicity, which is
originated from the division part of the global symmetry
$[SU(N_f)_L\times SU(N_f)_R\times U(1)_V]/[\mathbb Z_{N_c}\times(\mathbb
Z_{N_f})_V]$.  This violation is actually an anomaly when
$\operatorname{gcd}(N_c,N_F)\neq1$~\cite{Cordova2020a}.  As a result, at
least when $\operatorname{gcd}(N_c,N_f)\neq 1$, we find that there has
to be some topological quantum field theory on the domain wall of $\eta'$.  We further restrict the
effective theory by considering the large-$N_c$ argument by Witten
\cite{Witten1979}, which states that the part of the effective theory
only including glueballs are the same as that of the Yang-Mills theory.
The domain wall we consider corresponds to the change of $\eta'$ over
one period.  This corresponds to the shift of $\theta$ from $0$ to $2\pi
N_f$.
As we mentioned above, a topological quantum field theory on the interface of $\theta$ from $0$ to
$2\pi$ is $SU(N_c)_1$.  When $\theta$ changes from $0$ to $2\pi N_f$,
the theory $(SU(N_c)_1)^{N_f}$ can undergo a transition to another topological quantum field theory~\cite{Hsin2019}.  It has been suggested that the theory is $SU(N_c)_{N_f}$~\cite{Gaiotto2017}.  Due to the level-rank duality, an $SU(N_c)_{N_f}$
CS theory is identified with a $U(N_f)_{-N_c}$ CS theory.  Thus we can
expect that there is a $U(N_f)_{-N_c}$ CS theory on the domain wall.
In \cref{sec:dw_st}, we determine the effective theory on the pancake
when there is the background $U(N_f)_V/\mathbb Z_{N_c}$ gauge field.  We
find that if we couple the gauge field of the CS theory to the
background gauge field properly, then we can maintain the gauge
invariance of the background gauge field.  As a result, we confirm that
an excited state with a unit $U(1)$ charge belongs to the correct
representation of the global symmetry as a spin-$N_c/2$ baryon.  
In \cref{sec:vec}, we consider the coupling between the $SU(N_f)_L\times
SU(N_f)_R$ background gauge field and the domain wall.
Unlike the case in \cref{sec:dw_st}, the gauge field of the CS theory
cannot be used to recover the gauge invariance of the $SU(N_f)_L\times
SU(N_f)_R$ gauge field.
We propose that the vector mesons as the gauge field of the hidden local
symmetry should couple to the domain wall instead of the background
fields.  Due to the mass mixing of the vector mesons and the background
fields, the representation for a spin-$N_c/2$ baryon is correctly
reproduced.  Also, it is found that the vector meson cannot have nonzero
instanton charge for a consistency.  In \cref{sec:dual}, physical
implications of the coupling between the vector mesons and the domain
wall are explained.  The effective theory we derived provides us with
new understanding of the ``duality'' between the gluon and the vector
mesons.

\section{Practice: A domain wall and a monopole}
\label{sec:practice}

Before going into the technical detail, we start with a physical
question in QCD where the baryon number symmetry, $U(1)_B$, is weakly gauged. Let us consider a configuration where a magnetic monopole, whose magnetic charge under $U(1)_B$ is unity, is surrounded by the domain wall that connects $\eta' = 0$ and $\eta' = 2 \pi \sim 0$.
The WZW term of QCD contains
the following term:
\begin{align}
 \frac{i}{8 \pi^2}\frac{N_f}{N_c} \eta' dA_B dA_B,
\end{align}
where $A_B$ is the baryon gauge field where it is normalized such
that the baryon has the charge unity. This term induces the Witten
effect\footnote{
The modern understanding of the Witten effect associated with the axion domain wall has been discussed in Refs.~\cite{Hidaka2020a,Hidaka2020b}, where a higher-group structure is found among the higher form symmetries in the axion electrodynamics.
}~\cite{Witten1979c} when $\eta'$ changes from $0$ to $2 \pi$. The
electric charge, i.e., the baryon number, of this configuration is $N_f
/ N_c$. In the effective theory, such an object is not allowed by the
Dirac quantization condition for general values of $N_f$ and $N_c$.

This already provides an interesting puzzle. When we go back to QCD, the
monopole with the unit magnetic charge is allowed by the Dirac
quantization condition. For the first look it has a problem with the
quarks which has baryon number $1/N_c$, but it is actually consistent by
attaching the color magnetic flux to the monopole~\cite{Corrigan1976}.
For consistency, the effective theory should also allow such a monopole
to exist. But once it is allowed, since there is no gluon degrees of
freedom in the effective theory, one cannot attach the color magnetic
flux anymore, and the Dirac quantization condition seems to be just
violated.

In order to fix this problem, we need something topologically nontrivial on the domain wall.
The ``something'' should be either a term to cancel the Witten effect so
that the configuration is allowed or some term to make the configuration
impossible.
We will see that QCD choose the latter by having a CS theory on the
wall. 

\section{The effective theory for $\eta'$}
\label{sec:eta}

We consider large-$N_c$ QCD with $N_f$ massless Dirac fermions.
The (Euclidean) Lagrangian is
\begin{align}
  \mathcal L = \frac{N_c}{4\lambda^2}\tr(f\hodge f) +iN_c\bar\psi\slashed D\psi
  + i\theta \frac1{8\pi^2}\tr f^2.
\end{align}
At the leading order of the $1/N_c$ expansion, the $U(1)_A$ symmetry is recovered and we can write the effective action by using a $U(N_f )$-valued field $U$.
We decompose $U$ into $U(1)$ part and $SU(N_f)$ part as $U=e^{i\eta'}\widehat U$, $\widehat U\in SU(N_f)$, and focus on $\eta'$.
This decomposition is not determined uniquely, but if we fix the pion field $\widehat U$, this ambiguity is not relevant.

When we add small masses to quarks, the effective Lagrangian should include a term:
\begin{align}
  \frac{f_\pi^2\Lambda}{2}\tr(MU + M^\dag U^\dag),
\end{align}
where $M$ is the $N_f\times N_f$ mass matrix and $\Lambda$ is the scale of the strong interaction.
In the following, we discuss the effective theory of QCD in the massless limit.
The correction by the quark masses can always be taken into account by adding Eq.(3) as a small perturbation\footnote{
At $\theta = \pi$, with finite quark masses, $m$, the vacuum structure in the large $N_c$ limit actually depends on the order of two limits, $m \to 0$ and $1/N_c \to 0$ \cite{Gaiotto2018}. Our discussion of the massless quarks corresponds to $m \to 0$ first, where the $\eta'$ mass is larger than $m\Lambda$. However, the domain wall of our interest which connects $\eta' = \eta_0'$ and $\eta_0' + 2\pi$ always exists independent of the vacuum structure, and the discussion of the theories on the domain wall and on its boundary are unchanged even in the other limit since the effective theory is constructed only based on anomaly matching and the gauge invariance.
}.

It is known that, to the next to leading order of $1/N_c$ expansion, the
effective Lagrangian for $\eta'$ is
\begin{align}
  \mathcal L^{\mathrm{eff}}_{\eta'} = \frac{N_ff_\pi^2}{8}d\eta'\hodge d\eta' + \frac{f_\pi^2}{8N_f}m_{\eta'}^2\min_{n\in\mathbb Z}(N_f\eta'+\theta - 2\pi n)^2,
\label{eq:eta_eff}
\end{align}
where the $N_c$ dependence of the parameters is given as
\begin{align}
  f_\pi^2 = \mathcal O(N_c), \quad m_{\eta'}^2 = \mathcal O(1/N_c).
\end{align}
Here the normalization of $\eta'$ is chosen so that, under $U(1)_A$ transformation $\psi \rightarrow \exp(i\alpha\gamma^5)\psi$, it transforms as $\eta'\rightarrow \eta'+2\alpha$.
For this normalization, $\eta'$ is $2\pi$-periodic.
The coefficients are determined so as to reproduce the Witten-Veneziano formula~\cite{Witten1979,Veneziano1979}
\begin{align}
  m_{\eta'}^2 = \frac{4N_f}{f_\pi^2}\chi,
\end{align}
where $\chi$ is the topological susceptibility in the $SU(N_c)$ Yang-Mills theory.
The periodicity and $1/N_c$ expansion imply that the $\eta'$ potential
has a cusp at $N_f \eta' + \theta = n \pi$ with $n$ odd integers.

We consider how to couple the background gauge fields to the effective
theory.  The global symmetry of the theory is $[SU(N_f)_L\times
SU(N_f)_R\times U(1)_V]/[\mathbb Z_{N_c}\times (\mathbb
Z_{N_f})_V]$~\cite{Tanizaki2018}.  This is because the group with the
faithful action on the fermion field is $[SU(N_c)\times SU(N_f)_L\times
SU(N_f)_R\times U(1)_V]/[\mathbb Z_{N_c}\times (\mathbb Z_{N_f})_V]$
rather than the simple direct product $SU(N_c)\times SU(N_f)_L\times
SU(N_f)_R\times U(1)_V$.  In this section, we focus on the subgroup
\begin{align}
G_{\mathrm{sub}}:=[SU(N_f)_V\times U(1)_V]/[\mathbb Z_{N_c}\times(\mathbb Z_{N_f})_V],
\end{align}
and couple the corresponding background fields to QCD.

\subsection{Anomaly constraint}
Recently, it is shown that QCD has an anomaly involving
the $\theta$-periodicity and the global symmetry~\cite{Cordova2020a}.  We
can use this anomaly to restrict the coupling between the background
fields and $\eta'$ at large $N_c$.

In the following discussion, the division part of $G_{\mathrm{sub}}$
plays the main role.  In order to express the division part we should
introduce the 2-form background gauge fields for the $\mathbb
Z_{N_{c}}$ and $\mathbb Z_{N_{f}}$ groups (see Refs.~\cite{Gaiotto2017,Tanizaki2018}).  A 2-form $\mathbb Z_{N_{c}}$ gauge
field is realized as a pair of a 2-form $U(1)$ gauge field and a
1-form $U(1)$ gauge field,
$(B_c^{(2)},\widehat A_c)$, that
satisfies a constraint $N_cB_c^{(2)}=d\widehat A_c$, where the
normalization of $\widehat A_c$ is given as $\int_\Sigma d\widehat
A_c\in2\pi\mathbb Z$ for a closed 2-surface $\Sigma$.  They transform under a 1-form $U(1)$ gauge
transformation as
\begin{align}
  B_c^{(2)}\rightarrow B_c^{(2)} + d\lambda_c^{(1)},\quad \widehat A_c\rightarrow \widehat A_c + N_c\lambda^{(1)}_c,
  \label{1-form_c}
\end{align}
where $\lambda^{(1)}_c$ is an arbitrary 1-form $U(1)$ gauge field
whose normalization is given as $\int_\Sigma d\lambda^{(1)}_c\in2\pi\mathbb
Z$.  
As the gluon degrees of freedom, we introduce a dynamical $U(N_c)$ gauge
field $\widetilde a$ instead of the $SU(N_c)$ gauge field.  To eliminate
the extra degrees of freedom, we impose a constraint on the field
strength, $\widetilde f$, as
\begin{align}
\tr\widetilde f = d\widehat A_c.
\end{align}
Under the 1-form $U(1)$ gauge transformation in \cref{1-form_c},
$\widetilde a$ transforms as
\begin{align}
  \widetilde a \rightarrow \widetilde a+ \lambda_c^{(1)}\bm 1
\end{align}
to maintain the constraint.  The field strength $f$ for the $SU(N_c)$
gauge field in the action is replaced by
\begin{align}
  \widetilde f - B_c^{(2)}\bm 1.
\end{align}
Note that it is a 1-form gauge invariant $\mathfrak{su}(N_c)$-valued 2-form locally.

In the same way, we introduce a 2-form $\mathbb Z_{N_f}$ gauge field
$(B_f^{(2)},\widehat A_f)$, and a 1-form $U(N_f)$ gauge field
$\widetilde A_f$ that satisfy a constraint
\begin{align}
\tr\widetilde F_f = d\widehat A_f.
\end{align}
They transform under a 1-form gauge transformation as
\begin{align}
  \widetilde A_f \rightarrow \widetilde A_f + \lambda_f^{(1)}\bm 1,
  \quad
  B_f^{(2)} \rightarrow B_f^{(2)} + d\lambda_f^{(1)}
  ,\quad
  \widehat A_f \rightarrow \widehat A_f + N_f \lambda_f^{(1)}
  .
  \label{1-form_f}
\end{align}

Let $\widetilde A_V$ be the $U(1)_V$ gauge field, which transforms under
the 1-form gauge transformations as
\begin{align}
  \widetilde A_V \rightarrow \widetilde A_V - \lambda_c^{(1)} -\lambda_f^{(1)}.
  \label{1-form_V}
\end{align}
Using these fields, the covariant derivative is defined as
\begin{align}
  D\psi = (d -i \widetilde a -i \widetilde A_f -i \widetilde A_V)\psi,
\end{align}
which is invariant under the 1-form gauge transformations.
%
We define ``the instanton charge densities'' as 1-form gauge invariant
quantities,
\begin{align}
  &q_c := \frac1{8\pi^2}\left(\tr(\widetilde f^2) -\frac1{N_c}(d\widehat A_c)^2\right),
  \quad
  q_f := \frac1{8\pi^2}\left(\tr(\widetilde F_f^2) -\frac1{N_f}(d\widehat A_f)^2\right),
  \notag\\
  &q_V := \frac1{8\pi^2}\left(d\widetilde A_V +\frac1{N_c}d\widehat A_c+ \frac1{N_f}d\widehat A_f\right)^2
  .
\end{align}

In order for a $U(1)_A$ transformation $\psi\rightarrow\exp(i\alpha\gamma^5)\psi$ to compensate a constant shift of $\theta$, we introduce the counter term
\begin{align}
  \mathcal L_c = i\theta\frac1{N_f}(N_cq_f + N_cN_fq_V).
  \label{Lc}
\end{align}
Because a constant shift of $\theta$ is compensated by a shift of $\eta'$, we should add the following term to the effective Lagrangian:
\begin{align}
  \mathcal L_{\mathrm{topo}}^{\mathrm{eff}} = i(N_f\eta'+\theta)\frac1{N_f}(N_cq_f+N_cN_fq_V).
  \label{Lt}
\end{align}
Note that under the $U(1)_A$ transformation $\psi\rightarrow \exp(i\alpha\gamma^5)\psi$, $\eta'$ changes as $\eta'\rightarrow \eta'+2\alpha$.
However, as we will see in the following, this term alone added by
Eq.~\eqref{eq:eta_eff} cannot be the correct effective theory.
We need more terms in the effective Lagrangian.

Because of the division part of the symmetry, we can find that full
QCD has the following properties that should be maintained in the
effective theory:
\begin{itemize}
%

%

\item \textbf{Fractional instanton charges}

The instanton charges can have fractional values.
The values of the instanton charges are restricted as~\cite{Anber2019}
\begin{align}
  &Q_c:=\int q_c = -\frac{m_cm_c'}{N_c} + l_c
  ,\quad Q_f:=\int q_f = -\frac{m_fm_f'}{N_f} + l_f
  ,\notag\\ &Q_V:=\int q_V = \left(\frac{m_c}{N_c}+\frac{m_f}{N_f} + l_V\right)\left(\frac{m'_c}{N_c}+\frac{m'_f}{N_f} + l_V'\right)
  ,\notag\\ &m_{c,f},m_{c,f}' = 0,\ldots, N_{c,f}-1
  ,\quad l_c,l_f,l_V,l_V'\in\mathbb Z
  .
\end{align}
Note that $m_c,m_c',m_f,m_f',l_f,l_V,l_V'$ are determined by the background fields.
Only $l_c$ depends on the dynamical field.

\item \textbf{$Z(\theta)=0$ for some background fields}

For some background fields, the partition function $Z(\theta)$ becomes zero.
This is because the index of the Dirac operator is
\begin{align}
  \operatorname{ind}i\slashed D &= N_f Q_c + N_cQ_f + N_cN_fQ_V 
  \notag\\&= m_cm_f'+m_c'm_f + (N_fm_c+N_cm_f)l_V' + (N_fm_c'+N_cm_f')l_V \notag\\
  &\quad + N_c l_f  + N_f l_c + N_cN_f l_Vl_V'
  \notag\\&= l_{I} + N_f(m_cl_V'+m_c'l_V+N_cl_Vl_V') + N_fl_c 
  ,\notag\\ l_{I}&:= m_cm_f'+m_c'm_f +N_c(m_fl_V'+m_f'l_V + l_f)
  .
\end{align}
Again note that only $l_c$ depends on the dynamical fields and other
integers are determined by the external fields.  If $l_{I}$ is not an
integer multiple of $N_f$, the index is nonzero for all values of $l_c$,
and thus in this case the partition function is zero.
Conversely, when $l_{I}$ is an integer multiple of $N_f$ (for example $\widetilde A_f = \widehat A_f=0$), the partition function is in general non-vanishing.

\item \textbf{The anomaly between the $\theta$-periodicity and $G_{\mathrm{sub}}$}

Under the $2\pi$ shift of $\theta$, the unnormalized expectation value $\braket{X[\psi]}_{\theta}$ of some operator $X[\psi]$ transforms as
\begin{align}
  \braket{X[\psi]}_{\theta+2\pi} = \exp\left(2\pi i\frac{l_I}{N_f}\right)\braket{X[\psi]}_{\theta}, \label{x}
\end{align}
for the choice of the counter term \cref{Lc}.
This violation of the $2\pi$-periodicity of $\theta$ is actually anomaly if the greatest common divisor $L:=\operatorname{gcd}(N_c,N_f)$ is not equal to 1, because in this case this phase shift cannot be eliminated by adding a counter term to the Lagrangian as shown in Ref.~\cite{Cordova2020a}.

Because we choose the counter term so that a constant shift of $\theta$ is compensated by a $U(1)_A$ transformation, the $(\mathbb Z_{2N_f})_A$ transformation give the same phase shift as
\begin{align}
  \Braket{X\left[\exp\left(-i\frac{2\pi}{2N_f}\gamma^5\right)\psi\right]}_\theta = \exp\left(2\pi i\frac{l_{I}}{N_f}\right)\braket{X[\psi]}_\theta.
  \label{daa}
\end{align}
This means, when $l_I$ is not an integer multiple of $N_f$, a non-vanishing path integral requires an appropriate insertion of fermion operators such as $(\bar\psi\psi)^{l_I}$.
This is actually the anomaly between $(\mathbb Z_{N_f})_L$ and $G_{\mathrm{sub}}$~\cite{Tanizaki2018}.
Note that for a different choice of the counter term, the $2\pi$-shift of $\theta$ gives a different phase shift, and a constant $\theta$ shift cannot be compensated by a $U(1)_A$ transformation.
\end{itemize}

We can see that the effective Lagrangian in Eq.~(\ref{Lt}) added by
\cref{eq:eta_eff} is not satisfactory. By using the background with
fractional instanton charges, that should be allowed, the partition
functions and the $\theta$ dependence above do not quite match.
%
%
%
Let us take $l_{I}$ as an integer multiple of $N_f$. As we have seen,
the partition function, $Z(\theta)$, should not vanish in this
background.
In the effective theory, $Z(\theta)$ is obtained as the path integral
of $\eta'$.  By changing the variable as $\eta' \rightarrow \widetilde
\eta' = \eta' + 2\pi/N_f$, which leaves the kinetic and the mass terms
invariant, we obtain
\begin{align}
  Z(\theta) = e^{-2\pi i\frac{m_cm_c'}{N_c}}Z(\theta),
  \label{p1}
\end{align}
from Eq.~(\ref{Lt}).  When $m_cm_c'\neq 0$, the phase is not unity,
which implies $Z(\theta)=0$.

The property of the $2 \pi$ shift of $\theta$ is also not maintained.
Because of the term (\ref{Lc}), the $2\pi$-shift of $\theta$ gives the
phase $\exp(2\pi i(l_I/N_f + m_cm_c'/N_c))$, which is different from
\cref{x}.

In order to avoid these pathologies, we should add some term to the effective Lagrangian.
It is expected that the term contains a 4-form $C^{(4)}$ that depends only on the background fields and satisfies
\begin{align}
  \int C^{(4)} = \frac{m_cm_c'}{N_c} \mod 1
  . \label{condC}
\end{align}
For example, the term
\begin{align}
  -(N_f\eta'+\theta)C^{(4)}
\end{align}
seems to fix the pathologies.  However, we cannot write this type of
terms at least if $L:=\operatorname{gcd}(N_c,N_f)\neq1$.  This is
because, $C^{(4)}$ that satisfies the condition (\ref{condC}) cannot be
written as a field that only contains the background fields and is
invariant under the 1-form gauge transformations (\ref{1-form_c}),
(\ref{1-form_f}) and (\ref{1-form_V}) if $L\neq
1$~\cite{Cordova2020a}.  Since the original QCD Lagrangian is invariant
under the 1-form gauge transformation, the effective theory must be
invariant.

We show that $C^{(4)}$ cannot be written as a 1-form gauge invariant
field
by following Ref.~\cite{Cordova2020a}.  It is enough to show that, if
$L\neq1$, there are no solutions of the equation for variables $s$ and
$t$,
\begin{align}
 Q_f s + Q_V t =\frac{m_cm_c'}{N_c} \mod 1.
  \label{cfv}
\end{align}
Note that $s$ and $t$ do not depend on $m_c$, $m_c'$, $m_f$, $m_f'$,
$l_c$, $l_f$ or $l_V$, because these quantities only defined as the
integrals over the whole space, but $C^{(4)}$ is a local quantity.  In
order to eliminate $l_f$, $l_V$ and $l_V'$, the variable $s$ has to be
an integer and $t$ has to be an integer multiple of $N_cN_f/L$.  Then
the equation becomes
\begin{align}
  \frac{m_cm_c'}{N_c}\left(-1+\frac t{N_c}\right)
  +\frac{m_fm_f'}{N_f}\left(-s+\frac t{N_f}\right)
  +\frac{m_cm_f' + m_c'm_f}{N_cN_f}t
  = 0 \mod 1.
\end{align}
It follows from this that
\begin{align}
  -\frac 1{N_c} + \frac t{N_c^2} = R\in\mathbb Z
  ,\quad -\frac{s}{N_f} + \frac t{N_f^2} = P \in \mathbb Z
  ,\quad \frac{t}{N_cN_f} = J\in \mathbb Z.
  \label{RPJ}
\end{align}
The first and last equation imply
\begin{align}
  1 = N_f J- N_cR. \label{JR}
\end{align}
Because the right-hand side is an integer multiple of $L$, there is a
solution only if $L=1$.  Therefore, if $L\neq 1$ we cannot write
$C^{(4)}$ as a 1-form gauge invariant field.

Since the coupling between the background fields and $\eta'$ cannot
solve the problems, we need to extend the effective theory to include
more dynamical degrees of freedom. In the following we introduce a
dynamical domain wall and a CS theory on it.
%
%
In the potential for $\eta'$ in \cref{eq:eta_eff}, there are jumps between different branches.
In each branch labeled by an integer $n$, the potential for $\eta'$ is given as
\begin{align}
  \frac{f_\pi^2}{2}m_{\eta'}^2(N_f\eta'+\theta - 2\pi n)^2.
  \label{pot}
\end{align}
The jump between branches indicates that a dynamical domain wall (made of some heavy degrees of freedom) attaches at the location where $n$ changes.
The location is dynamically chosen such that the energy is minimized. 
We propose that in the effective action, there is the term
\begin{align}
  -2\pi i\int n C^{(4)}, 
  \label{kC}
\end{align}
where $n$ takes different values on the each side separated by the domain wall.
The action is accompanied by the term on the world volume,
\begin{align}
   i \int_{\partial X_4} c^{(3)},
\end{align}
where $X_4$ is the interior of the domain wall, and $c^{(3)}$ should be
some dynamical degree of freedom on the wall which is necessary to
maintain the gauge invariance.
The value of $n$ inside $X_4$ is one greater than the value outside $X_4$.

One can find that the problems are solved as follows.
If we change the variable in the path integral as $\eta'\rightarrow \widetilde{\eta}'=\eta'+2\pi/N_f$, we should also change $n$ as $n\rightarrow \widetilde n=n+1$ in order for the potential in \cref{pot} to be unchanged.
Due to the contribution from \cref{kC}, the unwanted phase in \cref{p1} is cancelled.
The phase shift \cref{x} is also reproduced in the same way.

The condition (\ref{condC}) does not determine the form of $C^{(4)}$ and $c^{(3)}$
uniquely, but the simplest possibility is to set
\begin{align}
   C^{(4)} = \frac{N_c}{8\pi^2}(B_c^{(2)})^2,
  \label{spt}
\end{align}
and introduce $c^{(3)}$ as the $U(1)_{-N_c}$ CS theory,
\begin{align}
  i\int_{\partial X_4}c^{(3)} = -i\frac{1}{4\pi}\int_{\partial X_4}(N_ccdc + 2cd\widehat{A}_c),
  \label{cs1}
\end{align}
where $c$ transforms as $c\rightarrow c-\lambda^{(1)}_c$ and is
normalized as $\int_\Sigma dc \in 2\pi\mathbb Z$.
The normalization condition is necessary for the CS action to be well-defined.
With this choice, the whole action is invariant under the 1-form transformation.
This provides a consistent effective theory.

Generally, if $C^{(4)}$ is given as \cref{spt}, a theory on the domain
wall has to be able to couple to the 2-form $\mathbb Z_{N_c}$ gauge
field $(B_c^{(2)},\widehat A_c)$, but it cannot be 1-form gauge
invariant.  This means that the theory has to have an anomaly for a
$\mathbb Z_{N_c}$ 1-form symmetry.  
In other words, the path integral of the theory on the wall is defined
by the combination with the four-dimensional symmetry protected topological
(SPT) phase (\ref{spt}), and it depends on how to take the
four-dimensional space $X_4$.  If the domain wall separates the
spacetime into two regions $X_4$ and $X_4'$, the difference between the
two choices is the integral over the whole space,
\begin{align}
  &-i\frac{N_c}{4\pi} \left(\int_{X_4}(B_c^{(2)})^2-\int_{X_4'}(B_c^{(2)})^2\right)\notag\\
  &=-i\frac{N_c}{4\pi}\int_{\text{whole space}} (B_c^{(2)})^2 = -2\pi i\frac{m_cm_c'}{N_c} \mod 2\pi i.
\end{align}
Thus, the dependence on the choice of $X_4$ is characterized by an element of $\mathbb Z_{N_c}$ in this case.

In the next subsection, we will see that \cref{spt} is the most plausible form of $C^{(4)}$, but logically, there are other possibilities of $C^{(4)}$, e.g., 
\begin{align}
  C^{(4)} =-\frac{R}{L} \frac{N_c^2}{8\pi^2}(B_c^{(2)})^2 - \frac{J}{L} \frac{N_cN_f}{4\pi^2}B_c^{(2)}(d\widetilde A_V + B_f^{(2)}) + \frac{JN_cN_f}{L}q_V + \frac{JN_c}{L}q_f,
  \label{sptl}
\end{align}
where $J$ and $R$ are integers that satisfy
\begin{align}
  JN_f - RN_c = L.
\end{align}
For this $C^{(4)}$, a theory on the domain wall has to have an anomaly that is controlled by the first two terms in \cref{sptl}.
Note that the last two terms in \cref{sptl} is 1-form gauge invariant.
The difference between two choices of the interior is calculated as
\begin{align}
  &-2\pi i\int\left[-\frac{R}{L} \frac{N_c^2}{8\pi^2}(B_c^{(2)})^2 - \frac{J}{L} \frac{N_cN_f}{4\pi^2}B_c^{(2)}(d\widetilde A_V + B_f^{(2)})\right] = -2\pi i\frac mL \mod 2\pi i,\notag\\
  &m:= -Rm_cm_c' - J(m_cm_f'+m_c'm_f).
\end{align}
Here $m$ can be any integer because $R$ and $J$ are relatively prime.
Thus we find that the theory on the wall has a $\mathbb Z_L$ anomaly for this $C^{(4)}$, which is different from the case (\ref{spt}) where a theory on the wall has a $\mathbb Z_{N_c}$ anomaly.
It is hard to believe that the theory on the domain wall depends on $L$, but we cannot deny this possibility only from the anomaly argument.
Especially, if $L=1$, we cannot show even the necessity of dynamical degrees of freedom on the wall.
In this case, $C^{(4)}$ can be written only using 1-form invariant background fields as
\begin{align}
  C^{(4)} = JN_cN_fq_V + JN_c q_f.
\end{align}
If $C^{(4)}$ is actually this form, there does not need to be dynamical degrees of freedom on the wall.

When we consider the situation that the $N_f\eta' + \theta$ changes rapidly from
$0$ to $2\pi k$, there are additional possibilities of the theory on the
wall other than the theory obtained by simply adding $k$ copies of the
same theory.  In this case, the theory has to have an anomaly controlled
by
\begin{align}
  -2\pi ik\int_{X_4} C^{(4)}.
\end{align}
For example, if $C^{(4)}$ is \cref{spt}, the theory on the wall can be
an $SU(N_c)_k$ CS theory rather than $(U(1)_{-N_c})^k$.

In summary, the effective action for the domain wall corresponding to $2\pi k$ shift of $N_f\eta'+\theta$ is
\begin{align}
  i\frac{2\pi k}{N_f}\int_{X_4}(N_cq_f + N_cN_fq_v - N_fC^{(4)}) + i\int_{\partial X_4}c_k^{(3)},
  \label{wall_theory}
\end{align}
where $X_4$ is an interior of the domain wall, $C^{(4)}$ is a
background 4-form field that satisfies the condition (\ref{condC}),
and $c_k^{(3)}$ is a dynamical 3-form field for which the theory
$i\int c_k^{(3)}$ has an anomaly controlled by the bulk action $-2\pi
ik\int C^{(4)}$.

\subsection{Additional large-$N_c$ argument}
Using a large-$N_c$ argument, we further restrict the effective action.
As discussed in Ref.~\cite{Witten1979}, at the leading order of the $1/N_c$ expansion, the terms in the effective action for QCD including only glueballs are the same as that of the Yang-Mills theory except for the replacement of $\theta$ by $N_f\eta'+\theta$.
Now we consider the effect of the background fields.
The only background fields that affect the gluon field $f$ is $B_c^{(2)}$.
If we focus on the gluon part, $B_c^{(2)}$ can be regarded as the background 2-form gauge field for the 1-form $\mathbb Z_{N_c}$ symmetry in the Yang-Mills theory.
It is known that, in the Yang-Mills theory with $\theta$ parameter, there is a domain wall where $\theta$ crosses $\pi$, and the effective theory on the domain wall has an anomaly controlled by \cref{condC} (if the spacetime is a spin manifold)~\cite{Gaiotto2017}.
Therefore we can conclude that \cref{condC} is the correct form of $C^{(4)}$ irrespective of $\operatorname{gcd}(N_c,N_f)$.
In addition, it is suggested in Ref.~\cite{Gaiotto2017} that, in the Yang-Mills theory, if $\theta$ changes rapidly from $0$ to $2\pi k$, the theory on the wall is $SU(N_c)_k$.
This expectation comes from an analogy with the $\mathcal N=1$ supersymmetric Yang-Mills theory.
Alternatively, by considering that the theory on the wall is $SU(N_c)_k$ in the UV, we can expect the same theory at low energy if we assume that the confinement in the bulk does not affect the theory on the wall.

According to this argument, we conclude that a theory on the domain wall corresponding to $2\pi k$ shift of $N_f\eta'+\theta$ is an $SU(N_c)_k$ CS theory.
Because of the level-rank duality, an $SU(N_c)_k$ CS theory is identified with a $U(k)_{-N_c}$ CS theory. 
Therefore, in the rest of this paper, we use
\begin{align}
  C^{(4)} = \frac{N_c}{8\pi^2}(B_c^{(2)})^2
  ,\quad
  c_k^{(3)} = -\frac{1}{4\pi}\left[N_c\tr\left(cdc-i\frac23c^3\right) + 2\tr(c)d\widehat{A}_c\right],
  \label{wall}
\end{align}
where $c$ is a 1-form $U(k)$ gauge field and transforms under the 1-form gauge transformation as 
\begin{align}
  c\rightarrow c - \lambda^{(1)}_c\bm 1.
\end{align}

\subsection{Resolution of the puzzle}
We can see the resolution of the puzzle in \cref{sec:practice}. A
monopole surrounded by the domain wall gives the integration of the
magnetic flux over a time slice $X_2$ of the wall as
\begin{align}
 \frac1{2\pi} \int_{X_2} \left(d \widetilde A_V+\frac1{N_c}d\widehat A_c+\frac1{N_f}\widehat A_f\right) = \frac1{N_c},
\end{align}
which means $\widehat A_c$ is non-vanishing on the wall.
By the Gauss law constraint on the wall from \cref{wall}, $dc-ic^2= - d\widehat A_c\bm 1/N_c$, $c$ is forced to have the magnetic flux,
\begin{align}
 \frac1{2\pi} \int_{X_2} d\tr(c) = -\frac{N_f}{2\pi N_c}\int_{X_2} d\widehat A_c = -\frac{N_f}{N_c}\mod 1,
\end{align}
which does not satisfy the quantization condition $\int_\Sigma d\tr(c)\in 2\pi\mathbb Z$.
Such a configuration is not allowed in accordance with the discussion of the Witten effect and the Dirac quantization.

\begin{figure}[t]
\centering
\includegraphics[width=0.25\hsize]{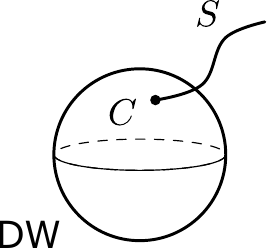}
\caption{A time slice of the domain wall with the operator (\ref{quark}). The surface $S$ is infinitely long or ended at another quark representing the color-electric flux tube.}
\label{fig:Wilson_loop_on_wall}
\end{figure}

The monopole can only be surrounded by the domain wall on which appropriate ``quark'' operators are inserted.
An example of such an operator is~\cite{Dierigl2014}
\begin{align}
  \exp\left(i\int_C\tr c\right)\exp\left(i\int_SB_c^{(2)}\right),
  \label{quark}
\end{align}
where $C$ is a contour on the domain wall, and $S$ is a two-dimensional surface whose boundary contains $C$.
The first part of the operator is the Wilson loop in a charge-$N_f$ representation of the $U(1)$ subgroup.
Let $C$ intersect with each time slice of the domain wall only once.
It is known that in the CS theory, the insertion of the Wilson loop causes the same effect as the singular gauge transformation~\cite{Moore1989}.
In this case, the operator corresponds to the gauge transformation by $g(\varphi)=\exp(i\varphi\bm 1/N_c)$ with an angular variable $\varphi$ around $C$.
Thus, when we insert the operator (\ref{quark}), the Gauss law constraint becomes
\begin{align}
  dc^g-i(c^g)^2 = -\frac1{N_c}d\widehat A_c\bm 1
  ,\notag\\
  c^g := gcg^{-1} + igdg^{-1}
  ,
\end{align}
where we only include the spatial components.
Due to the singularity, the integral of $d\tr(c^g)$ should be evaluated over $X_2\setminus D_{\varepsilon}$ with an infinitely small subregion $D_{\varepsilon}$ around $C$ as
\begin{align}
  \frac1{2\pi}\int_{X_2\setminus D_{\varepsilon}} d\tr(c^g) = -\frac1{2\pi}\int_{\partial D_{\varepsilon}} d\tr(igdg^{-1}) \mod 1= -\frac{N_f}{N_c} \mod 1.
\end{align}
Thus the Gauss law constraint can be satisfied.

The operator (\ref{quark}) itself does not have the baryon charge.
However, the whole object has the baryon charge $N_f/N_c$ due to the Witten effect.

In full QCD, the field $B_c^{(2)}$ in the definition of the operator
(\ref{quark}) represents the color-electric flux tube emerging from the
quark.  Because $C$ intersects with each time slice of the domain wall
only once, the surface $S$ has to be infinitely long or ended at another
quark.  See \cref{fig:Wilson_loop_on_wall}.
The color electric flux can also be understood as the one generated by the Witten effect since the monopole sources the color magnetic flux.

\section{The effective theory on the pancake}
\label{sec:dw_st}
The shift of $\eta'$ by $2\pi k/N_f$ for an integer $k$ can be reversed by shifting $U$ while keeping $\det U$ fixed, where $U$ is the $U(N_f)$-valued field to describe the pions and $\eta'$.
As a consequence, the domain wall corresponding to $2\pi k$ shift of $N_f\eta'+\theta$ could be bounded by a string.
At the core of the string, the chiral symmetry has to be recovered because otherwise the energy diverges there due to the kinetic term for $U$.

In this paper, we mainly consider domain walls corresponding to $2\pi
N_f$ shift of $N_f\eta'+\theta$ bounded by a string around which $\eta'$
has the monodromy $\eta'\rightarrow \eta'+2\pi$.  We call such an object
a pancake.  For this type of domain walls, the pion field does not need
to change because $\eta'=\eta'_0$ and $\eta'=\eta'_0+2\pi$ is the same
point in the target space of $U$.

In Ref.~\cite{Komargodski2018a}, it is proposed that a pancake
can be regarded as a baryon.  In this paper, we clarify how to couple
the background fields to pancakes and confirm that the objects
belong to an appropriate representation.

In the previous section, we have found that, in general, the effective theory on the domain wall depends on the choice of the interior as \cref{wall_theory}. 
We cannot consider that such a domain wall breaks and has a boundary.
However, when $k=N_f$, the wall theory does not depend on the choice of the interior because
\begin{align}
  2\pi i\int(N_cq_f+N_cN_fq_V-N_fC^{(4)})\in 2\pi i \mathbb Z,
\end{align}
where the integral is performed over the whole spacetime.  This consists
with the fact that $\eta'=0$ and $\eta'=2\pi$ represents the same vacuum
when $\theta=0$.  Then the action of the theory on the domain wall for
$k=N_f$ reduces to
\begin{align}
  i\frac{1}{4\pi}\int_{Y_3}\left[-N_c\tr\left(cdc-i\frac23 c^3\right) + N_c\tr\left(\mathcal A_fd\mathcal A_f -i\frac23 (\mathcal A_f)^3 \right)+ 2\left(\tr(\mathcal A_f) - \tr(c)\right)d\widehat A_c\right],
  \label{wall_theory_Nf}
\end{align}
where we define a $U(N_f)$ gauge field as
\begin{align}
  \mathcal A_f := \widetilde A_f +\widetilde A_V\bm 1.
\end{align}
Actually, this is the gauge field for the $U(N_f)_V$ global symmetry.
This field transforms under the 1-form gauge transformation as
\begin{align}
  \mathcal A_f \rightarrow \mathcal A_f - \lambda_c^{(1)}\bm 1.
\end{align}
We can check that this action is invariant under the zero-form and 1-form gauge transformations if $Y_3$ is a closed manifold.

For a domain wall bounded by strings, the action (\ref{wall_theory_Nf})
is not gauge invariant.  In order to recover the gauge invariance we
need to modify the action on the boundary.  Because $c$ is a dynamical
field only on the domain wall, we can impose a boundary condition.  On
the other hand, since $\mathcal A_f$ and $\widehat A_c$ are background
gauge fields living in the four dimensional spacetime, we cannot impose
a boundary condition.  A simple way to maintain the gauge invariance for
the background fields is to add the boundary term
\begin{align}
  -i\frac{1}{4\pi}\int_{\partial Y_3}N_c\tr(\mathcal A_fc).
  \label{boundary}
\end{align}
For this action, we can check the 1-form gauge invariance immediately.
To check the zero-form gauge invariance, we should consider a boundary condition for $c$.
Because a boundary condition should be imposed so that there are no boundary correction to the equation of motion, the condition is that one of the components of $c-\mathcal A_f$ is zero.
Note that this condition is 1-form gauge invariant.
To maintain the Lorentz invariance, the choice of a component that is imposed to be zero is restricted.
We decompose the gauge fields such as $c = c_tdt +\widetilde c$, and let $t$ denote such a component, i.e.,
we impose
\begin{align}
  c_t  - \mathcal A^f_t = 0.
\end{align}
This condition implies that $c$ transforms under the $U(N_f)_V$ transformation in the same way as $\mathcal A_f$ on the boundary.
Then we can check that the action is invariant under the zero-form gauge transformations.


Due to the boundary condition, the theory on the wall is equivalent to
the one with only including dynamical edge degrees of freedom, that is the
chiral version of the Wess-Zumino-Witten (WZW) model~\cite{Elitzur1989}.  In
the following, we set $\widehat A_c=0$.  By integrating out $c_t$ we
obtain the Gauss law constraint as
\begin{align}
  \widetilde f_c = 0.
  \label{gauss}
\end{align}
The solution of this constraint is
\begin{align}
  \widetilde{c} = iW\widetilde d W^{-1},\quad W\in U(N_f).
\end{align}
Then the action reduces to
\begin{align}
  S_p &= i\frac{N_c}{4\pi}\int_{\partial Y_3}\tr\left( W\widetilde d W^{-1} W\partial_t W^{-1}dt +2iW\widetilde dW^{-1} \mathcal A^f_tdt - \widetilde{\mathcal A}_f\mathcal A^f_tdt\right)
  \notag\\
  &\quad + i\frac{N_c}{4\pi}\int_{Y_3}\tr\left(-\frac13 (WdW^{-1})^3 + \mathcal A_fd\mathcal A_f -i\frac23\mathcal A_{f}^{3}\right).
  \label{edge}
\end{align}
Note that under the $U(N_f)_V$ gauge transformation, the fields transform as
\begin{align}
  \mathcal A_f\rightarrow g_f\mathcal A_fg_f^{-1}+ig_fdg_f^{-1},\quad W \rightarrow g_fW
  ,\quad
  g_f\in U(N_f)_V,
\end{align}
on the boundary.

We can read off the baryon charge $B$ from the action (\ref{edge}),
\begin{align}
  B = \frac{1}{2\pi}\int_{\partial Y_2} \tr(\mathcal A_f-iW dW^{-1}) - \frac1{2\pi}\int_{Y_2} \tr(\mathcal F_f),
  \label{baryon}
\end{align}
where $Y_2$ is the slice of $Y_3$ at a fixed $t$.
There is no contribution from the external fields when we take a gauge such that no Dirac string penetrates the pancake due to the cancellation between the bulk and boundary contributions. 
Therefore the baryon number is the winding number of $W$ around the slice corresponding to $\pi_1(U(N_F))=\mathbb Z$.

It has been proposed that the operator $\rho(W)$ for the $\operatorname{Sym}^{N_c}(\Box)$ representation $\rho$ describes a baryon with spin $N_c/2$~\cite{Komargodski2018a}.
This operator actually belongs to the correct representation of spin $N_c/2$ baryons because under the $U(N_f)_V$ transformation, it transforms
\begin{align}
  \rho(W) \rightarrow
  \rho(g_f)\rho(W),
\end{align}
which means that the corresponding state has baryon number one, and belongs to the representation $\operatorname{Sym}^{N_c}(\Box)$ of $SU(N_f)_V$.


\section{The role of the vector mesons}
\label{sec:vec} 

Now we obtain the consistent effective theory and boundary conditions on
the wall. We have seen that we obtained the consistent picture by
coupling $c$ to the external field ${\cal A}_f$ on the boundary. Here,
we try to recover all the background fields $A_L$ and $A_R$ for the
$SU(N_f)_L$ and $SU(N_f)_R$ symmetry. 
Once they are recovered, the vectorial part of the external field $\mathcal A_f$ does no longer transform properly under the full global symmetry, and thus the effective action in \cref{wall_theory_Nf,boundary} is not invariant under the $SU(N_f)_{L,R}$ gauge transformations. 
This is a problem since this breaks the anomaly matching condition for $SU(N_f)_L\times SU(N_f)_R$.
One may use the Nambu-Goldstone field $U$ to compensate the gauge transformation, but it is not possible on the boundary, since the chiral symmetry is unbroken there.
We propose to include the vector meson in the theory so that the appropriate boundary condition can be imposed.

The background gauge field for $SU(N_f)_{L,R}$ is introduced via
$U(N_f)$ valued gauge fields $\widetilde A_{L,R}$ in the same way as
$SU(N_f)_V$ gauge field.  We impose constraints
\begin{align}
  \tr\widetilde F_{L,R} = d\widehat A_f,
\end{align}
and the fields transform under the 1-form gauge transformation as
\begin{align}
  \widetilde A_{L,R} \rightarrow \widetilde A_{L,R} + \lambda^{(1)}_f\bm 1.
\end{align}
Similarly to the $SU(N_f)_V$ representation, they couple to the domain wall via $N_f\eta'+\theta$ as\footnote{
The counter term is chosen so that a $U(1)_A$ transformation compensates a constant shift of $\theta$ as
\begin{align}
  \mathcal L_c = i\theta\frac{N_c}{8\pi^2}(N_cq_L/2+N_cq_R/2+N_cN_fq_V).
\end{align}
This is because, under the $U(1)_A$ transformation $\psi \rightarrow \exp(i\alpha\gamma^5)\psi$, the Lagrangian is shifted as
\begin{align}
  \mathcal L \rightarrow \mathcal L +i\alpha(2N_fq_c + N_cq_L + N_cq_R + 2N_cN_fq_V).
  \label{c}
\end{align}
This shift is obtained as follows.
If the massless Dirac fermions $\psi$ couple to the gauge fields $\mathscr L$ and $\mathscr R$ as $(d-i\mathscr LP_L -i\mathscr RP_R)\psi$ with the projection operators $P_{L,R}$, the anomaly is characterized by the 6-dimensional action
\begin{align}
  S_6 =\frac{2\pi}{3!(2\pi^3)}\int\tr[(d\mathscr L-i\mathscr L^2)^3 - (d\mathscr R-i\mathscr R^2)^3].
\end{align}
We set $\mathscr L = \widetilde a + \widetilde A_L +\widetilde A_V\bm 1 + A_A\bm1$ and $\mathscr R=\widetilde a + \widetilde A_R + \widetilde A_V \bm1- A_A\bm1$.
The linear term in $A_A$ is given as the surface integral
\begin{align}
  S_6\supset \frac1{8\pi^2}\int [A_A(2N_fq_c + N_cq_L + N_c q_R + 2N_cN_f q_V].
\end{align}
This means that under the $U(1)_A$ transformation, the Lagrangian changes as \cref{c} if $A_A=0$.
}
\begin{align}
  &i(N_f\eta' + \theta)\frac1{N_f}\left(\frac12N_c q_L + \frac12N_c q_R + N_cN_fq_V\right)
  \notag\\
  &= i(N_f\eta'+\theta)\frac{1}{4\pi N_f}\left[\frac12N_c\tr(\mathcal F_L^2) + \frac12N_c\tr(\mathcal F_R^2)
  + (d\tr(\mathcal A_L)+d\tr(\mathcal A_R))d\widehat A_c \right],
  \label{etaLR}
\end{align}
where we define
\begin{align}
  &q_{L,R} := \frac1{8\pi^2}\left(\tr(\widetilde F_{L,R}^2)-\frac1{N_f}(d\widehat A_f)^2\right),
  \quad
  \mathcal A_{L,R} := \widetilde A_{L,R} + \widetilde A_V,
  \quad
  \mathcal F_{L,R}:= d\mathcal A_{L,R}-i\mathcal A_{L,R}^2.
\end{align}
This gives the theory on a domain wall for $k=N_f$ as\footnote{
The domain wall action depends on the interior if $N_c\int(\tr(\mathcal{F}_L^2)/2+\tr(\mathcal{F}_R^2)/2)/(8\pi^2)$ is not an integer.
To avoid this, we consider $\mathcal A_{L,R}$ with the same instanton numbers.
If we couple $\mathcal A_L$ and $\mathcal A_R$ with different instanton charges among them, $U$ becomes singular at a point as we discuss later.
We need to modify the effective theory to describe such a point.
}
\begin{align}
  & i\frac{1}{4\pi}\int_{Y_3}\biggl[-N_c\tr\left(cdc-i\frac23 c^3\right) + 2\left(\tr(\mathcal A_L)/2 + \tr(\mathcal A_R)/2 - \tr(c)\right)d\widehat A_c 
  \notag\\
  &\phantom{i\frac{1}{4\pi}\int_{Y_3}\biggl[}+ \frac12N_c\tr\left(\mathcal A_Ld\mathcal A_L -i\frac23 (\mathcal A_L)^3\right)
  + \frac12N_c\tr\left(\mathcal A_Rd\mathcal A_R -i\frac23 (\mathcal A_R)^3\right) \biggr].
  \label{wall_theory_LR}
\end{align}
However, for $Y_3$ with a boundary, we cannot recover the gauge invariance in the same way as in the previous section because there are two independent background fields $\mathcal A_L$ and $\mathcal A_R$.
Although it seems possible that this problem is solved by using $U$ and replacing $\mathcal A_R$ by $U\mathcal A_RU^{-1}+iUdU^{-1}$,
we cannot use $U$ on the boundary because the chiral symmetry is
restored there.  Our claim is that, to solve this problem, we should use
the vector mesons as the gauge field for the hidden local symmetry.

\subsection{Hidden Local symmetry}

In the formulation of the hidden local symmetry, the effective Lagrangian of the pions and the vector mesons is written in terms of a gauge theory as follows:
\begin{align}
  &\mathcal L_h = \mathcal L_V + a\mathcal L_A
  ,\notag\\
  &\mathcal L_V = -\frac{f_\pi^2}4\tr(D_\mu\xi_L\xi_L^\dag + D_\mu\xi_R\xi_R^\dag)^2 
  ,\notag\\
  &\mathcal L_A = -\frac{f_\pi^2}4\tr(D_\mu\xi_L\xi_L^\dag - D_\mu\xi_R\xi_R^\dag)^2
  ,\notag\\
  &D\xi_{L,R} = d\xi_{L,R} -iv\xi_{L,R} + i\xi_{L,R}\mathcal A_{L,R}
  ,
  \label{L_v}
\end{align}
where $v$ is the gauge field for the hidden local symmetry $U(N_f)_h$, and $\xi_L$ and $\xi_R$ relate to the $U(N_f)$-valued field $U$ that involves the pion field and $\eta'$ as 
\begin{align}
  \xi_L^\dag\xi_R = U.
\end{align}
Under the gauge transformations, the fields transform as
\begin{align}
  &v\rightarrow hvh^{-1}+ihdh^{-1},
  \quad
  \mathcal A_{L,R}\rightarrow g_{L,R}\mathcal A_{L,R}g_{L,R}^{-1} + i g_{L,R}dg_{L,R}^{-1},
  \quad
  \xi_{L,R} \rightarrow h\xi_Lg_{L,R}^{-1},
  \notag\\
  &h\in U(N_f)_h,
   \quad
  g_{L,R}\in U(N_f)_{L,R}.
\end{align}
To maintain the 1-form gauge invariance, $v$ transforms as
\begin{align}
  v \rightarrow v - \lambda^{(1)}_c\bm 1.
\end{align}
The term $\mathcal L_A$ does not depend on $v$ and reduces to the kinetic term for $U$.
The term $\mathcal L_V$ is regarded as the mass term for $v$.
Due to this term, fields with the $U(N_f)_h$ charge also couple to $A_L$ and $A_R$.
Actually, when we fix the gauge so that $\xi_L^\dag=\xi_R=:\xi$, to maintain this gauge under the global transformation, fields with the $U(N_f)_h$ charge are transformed by $h(\xi,g_L,g_R)$ that satisfies
\begin{align}
  g_L\xi h^{-1}(\xi,g_L,g_R) = h(\xi,g_L,g_R)\xi g_R^{-1}.
\end{align}
Therefore if $g_L=g_R=:g_f$, then $h(\xi,g_L,g_R)=g_f$, which means that fields with the $U(N_f)_h$ charge has the corresponding $U(N_f)_V$ charge.
The effective theory including $v$ should reduce to the effective theory for $U$ if the mass term for $v$ is minimized, i.e., if it is satisfied that
\begin{align}
  &v = \frac12(\mathcal{A}_L^{\xi_L} + \mathcal{A}_R^{\xi_R}),\notag\\
  &\mathcal{A}_{L,R}^{\xi_{L,R}}:=\xi_{L,R}\mathcal{A}_{L,R}\xi^{-1}_{L,R} +i\xi_{L,R}d\xi_{L,R}^{-1}.
  \label{mass}
\end{align}

\subsection{Coupling between the pancake and the vector mesons}
We propose that, the theory on the pancake should be written using the vector mesons $v$ as
\begin{align}
  &
  i\frac{1}{4\pi}\int_{Y_3}\left[-N_c\tr\left(cdc -i\frac23 c^{3}\right) + N_c\tr\left(vdv -i\frac23 v^3\right) + 2(\tr(v)-\tr(c))d\widehat A_c\right]
  \notag\\&
  +i\frac{N_c}{4\pi}\int_{\partial Y_3}\tr(vc)
  ,
  \label{wall_v}
\end{align}
up to terms we discuss in \cref{sec:c}\footnote{
When we consider the pancake configuration made of a subset of flavors, i.e., the $U(n_f)$ subgroup of $U(N_f)$, we can start from the hidden local symmetry only for the subgroup $U(n_f)$ and ignore the rest of NG bosons and vector mesons.
Then the gauge group of $c$ and $v$ are both reduced to $U(n_f)$ while the CS levels $-N_c$ are unchanged.
}.
For this action, the gauge invariance of the effective theory on the
pancake is maintained in the same way as in \cref{sec:dw_st}.  In order
for the CS term for $v$ to be included, $\tr(f_v^2)$ must couple to
$\eta'$, and the coupling (\ref{etaLR}) has to be reproduced when we
minimize the mass term for $v$, i.e., by substituting Eq.~(\ref{mass}).
Therefore the bulk action contains the coupling of $v$ to $\eta'$ and
$\theta$ as follows:
\begin{align}
  &i\eta'N_cq_v + i\theta\left(bN_cq_v + (1-b)\frac1{N_f}(N_cq_L/2+N_cq_R/2+N_cN_fq_V)\right),
 \notag\\ 
  &q_v := \frac{N_c}{8\pi^2}\tr\left[\left(f_v + \frac1{N_c}d\widehat A_c\bm 1\right)^2\right],
  \label{eta_v}
\end{align}
where the parameter $b$ is left arbitrary within our discussion.
We can check that this reproduces \cref{etaLR}.
When the mass term is minimized, it is satisfied that
\begin{align}
  &\tr(f_v^{\,2}) = \frac12\tr(\mathcal F_L^2) + \frac12\tr(\mathcal F_R^2) -\frac12 d\tr[(\mathcal A_L^{\xi_L} -\mathcal A_R^{\xi_R})(\xi_L\mathcal F_L\xi_L^{-1}-\xi_R\mathcal F_R\xi_R^{-1})],
  \notag\\
  &\tr(f_v) = N_fd\widetilde A_V + d\widehat A_f.
\end{align}
By substituting this, we obtain \cref{etaLR} added by the term
\begin{align}
  i\frac{N_c}{16\pi^2}(\eta'+b\theta) d\tr[(\mathcal A_L^{\xi_L} -\mathcal A_R^{\xi_R})(\xi_L\mathcal F_L\xi_L^{-1}-\xi_R\mathcal F_R\xi_R^{-1})].
\end{align}
This term is admissible because $\eta'$ couples to the exterior
derivative of a locally gauge invariant field, and thus the term is
invariant under constant shifts of $\eta'$.  Such a term does not change
the property that constant shifts of $\theta$ are compensated by shifts
of $\eta'$, which we used to determine the coupling of $\eta'$ to the
background fields.

It is not enough that \cref{etaLR} is reproduced when the mass term is minimized.
Moreover, we need to check that the anomaly between $(\mathbb Z_{2N_f})_A$ and $G_{\mathrm{sub}}$, \cref{daa}, is reproduced.
Under the $(\mathbb Z_{2N_f})_A$ transformation, $\eta'$ is shifted as $\eta'\rightarrow \eta'+2\pi /N_f$.
The phase shift is reproduced if and only if the instanton number for $v$ satisfies
\begin{align}
  \frac{1}{8\pi^2}\int\left(N_cq_v -N_cq_L/2-N_cq_R/2-N_cN_fq_V \right)\in N_f\mathbb Z
  \label{cond_v}
\end{align}
This means that if there are no background fields, the instanton charge for $v$ is restricted as an integer multiple of $N_f$.
The simplest possibility is that $v$ cannot have instantons on their own.

From the low-energy point of view, the left-hand side of \cref{cond_v}
is always zero for the following reason.  For simplicity, we set
$\widehat A_c=0$.
If the background fields $\mathcal A_{L,R}$ have nonzero
instanton charge, there should be topologically nontrivial transition
functions $g_{L,R}$ of the $U(N_f)_{L,R}$ transformation.
Since the instanton charge is gauge invariant, the instanton charge for $v^{\xi_{L,R}}:=\xi_{L,R}v\xi_{L,R}^{-1} + i\xi_{L,R}d\xi_{L,R}^{-1}$ is the same as that for $v$.
Since $v^{\xi_{L,R}}$ gauge transform in the same way as $\mathcal
A_{L,R}$, they have the same transition function $g_{L,R}$ and thus the same instanton charge as $\mathcal A_{L,R}$.
This means the left-hand side of \cref{cond_v} is
zero\footnote{
This also means that the instanton charge for $\mathcal A_L$ and $\mathcal A_R$ is the same as long as we assume that $\xi_L$ and $\xi_R$ are non-singular.
}.

The condition (\ref{cond_v}) has to be maintained as long as the
effective theory makes sense such as at finite temperatures.  One thing
we already have assumed is that the full theory admits a string around
which $\eta'$ winds by restoring the chiral symmetry at the core.  Here
we additionally need to assume that the effective theory does not admit
instantons for $v$.  This assumption seems to prohibit a simple UV
completion of the hidden local symmetry by embedding $\xi_L$ and $\xi_R$
into scalar fields whose target space is linear.  In such a UV theory,
it is expected that the mass term for $v$ can vanish, and thus the
singular points of $\xi_{L,R}$ are allowed, which means there are
configurations with non-zero instanton number for $v$ even if there are
no instanton solutions.


Another thing we should check is that whether the domain wall can be identified with a spin-$N_c/2$ baryon.
From the point of view of the effective theory on the domain wall, $v$ should be regarded as a background field although it is dynamical in the four-dimensional spacetime.
The field $\rho(W)$ on the edge, which is defined similarly in the previous section, transforms under the $\operatorname{Sym}^{N_c}(\Box)$ representation of $U(N_f)_h$.
In the four-dimensional theory, the pancake where the edge state corresponding to $\rho(W)$ excites has the correct charge of $U(N_f)_V$ due to the mass term for $v$. 
Thus the object is regarded as a spin-$N_c/2$ baryon correctly.
The baryon number is given by \cref{baryon} with the replacement of $\mathcal A_f$ by $v$,
\begin{align}
  B = \frac{1}{2\pi}\int_{\partial Y_2} \tr(v-iW\widetilde dW^{-1}) - \frac1{2\pi}\int_{Y_2} \tr(f_v),
  \label{baryon_v}
\end{align}
The contribution from $v$ is again cancelled as long as we take a gauge where a Dirac string does not penetrate the pancake.

The coupling of $\theta$ to the vector mesons and background fields in
\cref{eta_v} affects what happens if we insert an interface of $\theta$.
Even if $b$ is not an integer, the effective theory on the interface
from $0$ to $2\pi N_f$ does not depend on the choice of the interior.
This is because by using $\xi_L$ and $\xi_R$, we can rewrite the term
linear in $b$ in \cref{eta_v} as
\begin{align}
  &ib\theta\left(N_cq_v -\frac1{N_f}(N_cq_L/2 + N_cq_R/2 + N_cN_f q_V)\right)
  \notag\\
  &=
  i\frac{b\theta}2\sum_{H=\{L,R\}} \frac1{4\pi}d\biggl(N_c\tr\left[(v-\mathcal A_H^{\xi_H})(f_v+\xi_H\mathcal{F}_{H}\xi_H^{-1}) + i\frac13(v-\mathcal A_H^{\xi_H})^3 \right] 
  + 2\tr(v-\mathcal A_H^{\xi_H})d\widehat A_c\biggr)
  ,
\end{align}
where we have used
\begin{align}
  &\tr(f_v^2)-\tr(\mathcal F_{L,R}^2) = d\tr\left[(v-\mathcal A_{L,R}^{\xi_{L,R}})(f_v+\xi_{L,R}\mathcal F_{L,R}\xi_{L,R}^{-1}) + i\frac13(v-\mathcal A_{L,R}^{\xi_{L,R}})^3\right].
\end{align}
Here since the exterior derivative acts on the locally gauge invariant quantity, we can safely use the Stokes theorem.
Thus the effective action on the interface is the sum of \cref{wall_theory_LR} and
\begin{align}
  i\frac{b}{2}\sum_{H=\{L,R\}}\frac1{4\pi}\int_{V_3}\left(N_c\tr\left[(v-\mathcal A_H^{\xi_H})(f_v+\xi_H\mathcal F_{H}\xi_H^{-1}) + i\frac13(v-\mathcal A_H^{\xi_H})^3 \right] + 2\tr(v-\mathcal A_H^{\xi_H})d\widehat A_c\right)
  .
  \label{theta_v}
\end{align}

We can consider the configuration that the pancake attaches to the
interface.  This happens depending on how fast $\theta$ varies on the
interface.  On the edge of the pancake, we cannot use $\xi_L$ and
$\xi_R$ because the chiral symmetry is restored there.  This means that
near the core of the string where the order parameter of the chiral symmetry breaking,
$f_\pi$, approaches to zero, the value of $b$ should approach to an
integer, $m$,
\begin{align}
 b \to m.
\end{align}
The contribution from the integer part, $m$, of \cref{theta_v} does not
depend on $\xi_L$ and $\xi_R$, and is given by
\begin{align}
  i\frac{m}{4\pi}\int_{V_3}&\left[N_c\tr\left(vdv-i\frac23v^3\right)+ 2\bigl(\tr(v)-\tr(\mathcal A_L)/2-\tr(\mathcal A_R)/2\bigr)d\widehat A_c
  \right.
  \notag\\
  &\left.
 -\frac12N_c\tr\left(\mathcal A_Ld\mathcal A_L-i\frac23\mathcal A_L^3\right) -\frac12N_c \tr\left(\mathcal A_Rd\mathcal A_R-i\frac23\mathcal A_R^3\right) \right]
  .
  \label{n}
\end{align}
In this sense, we can say that on the interface, there is a CS term for $v$ with the level $mN_c$.
The following two values of $b$ are rather special:
\begin{itemize}
\item $b=0$\\
The vector mesons $v$ do not couple to the interface.
When the pancake attaches to the interface, there is only the CS term for $v$ on the pancake and not those for $c$ and the background fields.
On the other region of the interface, there are the CS terms for $c$ and
      the background fields, and not that for $v$.
\item $b=1$\\
The vector mesons $v$ couple to the interface via the CS term with the
      level $N_c$.
When the pancake attaches to the interface, there are no fields on the pancake except for the edge.
On the other region of the interface, there are the CS terms for $c$ and $v$, and not those for the background fields.
\end{itemize}


To summarize, we consider the pancake and the interface for $N_f=1$.
For simplicity, we set $\widehat A_c=0$, and only consider the $\omega$ meson field, the $U(1)$ gauge field $c$ of the CS theory, and the $U(1)_V$ background gauge field $A_V$.
We set $\xi_{L,R}=\exp(i\phi_{L,R})$.
The coupling of these fields to $\eta'$ and $\theta$ is
\begin{align}
  N_c\eta'd\omega^2 +N_c\theta (b\,d\omega^2 + (1-b)dA_V^2).
\end{align}
The theory on the pancake is
\begin{align}
  i\frac{N_c}{4\pi}\int_{Y_3}(-cdc+\omega d\omega) + i\frac{N_c}{4\pi}\int_{\partial Y_3}\omega c.
\end{align}
The theory on the interface is
\begin{align}
  &i\frac{N_c}{4\pi}\int_{X_3}(b(\omega - A_V+d\phi_L/2+d\phi_R/2)(d\omega + dA_V) -cdc+ A_VdA_V)\notag\\
  &= \begin{dcases}
  i\frac{N_c}{4\pi}\int_{X_3}(-cdc+A_VdA_V) &\text{for }b=0,\\
  i\frac{N_c}{4\pi}\int_{X_3}(-cdc+\omega d\omega) &\text{for }b=1.
  \end{dcases}
\end{align}

\subsection{\label{sec:c}Relation to the generalized WZW term with the vector mesons}
In the context of the hidden local symmetry, the possible generalization
of WZW term that includes the vector mesons has been
discussed~\cite{Fujiwara1985}.  We consider the relation between our
proposal of the coupling between $\eta'$ and $v$ in \cref{eta_v}, and
the generalized WZW term.  In this subsection, we only consider the case
where $\widehat A_c=0$, which means the 1-form symmetry corresponding
to $\lambda_c^{(1)}$ is fixed.  Especially, $\mathcal A_{L,R}$ are
1-form gauge invariant.  Note that we consider the WZW term
corresponding to the anomaly for $U(N_f)_L\times U(N_f)_R$ because the
axial $U(1)$ symmetry is restored at the leading order of the $1/N_c$
expansion. 
The external gauge field for the axial symmetry is later taken to be
vanishing, $A_A=0$, since the symmetry is broken when we include the
$1/N_c$ corrections in the effective Lagrangian.

We can only add the gauge invariant terms to the WZW term in order to
match the 't~Hooft anomaly, and thus possible additional terms consist
of the gauge covariant building blocks $f_v$, $\widehat{\mathcal
F}_{L,R}:=\xi_{L,R}\mathcal F_{L,R}\xi_{L,R}^{-1}$ and
$\widehat\alpha_{L,R}:=\mathcal A_{L,R}^{\xi_{L,R}}-v$.  The additional
term to the WZW term is written as a linear combination of four terms
as~\cite{Furui1986}
\begin{align}
  &\Gamma[\xi_L,\xi_R,v,\mathcal A_L,\mathcal A_R] = \Gamma_{\mathrm{WZW}}[\xi_L^\dag\xi_R,\mathcal A_L,\mathcal A_R] + \Gamma_v ,\notag\\
  &\Gamma_v:= -i\frac{N_c}{16\pi^2}\int_{M^4}\sum_{i=1}^4c_i\mathcal L_i
  ,\notag\\
  &\mathcal L_1 = \tr(\widehat \alpha_L^3\widehat\alpha_R - \widehat\alpha_R^3\alpha_L)
  ,\notag\\
  &\mathcal L_2 = \tr(\widehat\alpha_L\widehat\alpha_R\widehat\alpha_L\widehat\alpha_R)
  ,\notag\\
  &\mathcal L_3 = \tr(f_v(\widehat\alpha_L\widehat\alpha_R-\widehat\alpha_R\widehat\alpha_L))
  ,\notag\\
  &\mathcal L_4 = \frac12\tr\left(\widehat{\mathcal F}_L(\widehat\alpha_L\widehat\alpha_R-\widehat\alpha_R\widehat\alpha_L)-\widehat{\mathcal F}_R(\widehat\alpha_R\widehat\alpha_L-\widehat\alpha_L\widehat\alpha_R)\right)
  .
  \label{wz_vec}
\end{align}
The four terms, $\mathcal L_i$, are the general gauge invariant 4-forms
that conserve parity and charge conjugation but violate the intrinsic
parity.

The term (\ref{eta_v}) is included if we choose
$c_1=2/3,c_2=-1/3,c_3=1,c_4=1$ as\footnote{ The same values of $c_i$ are
proposed in Ref.~\cite{Karasik2020a} in the similar context.  }
\begin{align}
  \Gamma \supset i\frac{N_c}{8\pi^2}\int\eta'\tr(f_v^2).
  \label{c_i-v}
\end{align}
However, this does not mean that the values of $c_1,\ldots,c_4$ are
fixed for the following reason.  We can add freely terms in the form
$d\eta'A^{(3)}$ for a locally gauge invariant 3-form field $A^{(3)}$ because they
only add locally gauge invariant terms to the effective action on the
wall.  The problem is how to deal with them on the edge of the domain
wall if $A^{(3)}$ includes $\xi_{L,R}$.  On the edge, the chiral
symmetry is restored and we cannot use $\xi_{L,R}$.  A simple
possibility
is that they
become zero at the point where the chiral symmetry is restored.
By considering the finite thickness of the string, it is possible that the coefficients $c_i$ gradually deform into a required combination as we approach to the core of the string.
If the terms behaves like that, we can add them to the effective theory without
affecting the argument in the previous section.

\section{Summary and discussion}
\label{sec:dual} 

By the consideration of the large $N_c$ limit of QCD, the $\eta'$
potential should have cusps at $N_f \eta' + \theta = n \pi$ with $n$
odd integers. We find that the consistent effective theory of the
$\eta'$ meson should be accompanied by the dynamical domain wall which
appears when $N_f \eta' + \theta$ goes across the cusp. On the domain
wall, there should be a Chern-Simons (CS) theory in order for the anomaly matching to
be satisfied.

The domain wall can have an edge where the chiral symmetry is
restored. In this circumstance, in order to impose a consistent boundary
condition, we find that the gauge boson of the hidden local symmetry,
i.e., the vector mesons, should couple to that of the CS theory. The
whole framework gets consistent by this construction, and the quantum
numbers of the pancake excitations are correctly obtained as the
spin-$N_c/2$ baryons.

It is interesting to find that the existence of the pancake
configuration is made possible by the presence of the vector mesons, and
thus it is essential to include them in the effective theory.

Another interesting finding is that the consistency of the periodicity
of $\eta'$ requires that the vector mesons should not have a
non-vanishing instanton number. This indicates that a simple UV
completion of the hidden local symmetry by a linear sigma model is not possible. This has
some implication for the discussion of the finite temperature system or
in general for the chiral phase transition.
The vector mesons should not simply go to massless near the transition
point.

By the analogy with the Seiberg duality~\cite{Seiberg1995} in
supersymmetric QCD, the effective theory as a $U(N_f)$ gauge theory
based on the hidden local symmetry can be interpreted as some ``dual''
description of QCD~\cite{Seiberg2001, Harada1999, Komargodski2011,
Kitano2011, Abel2012, Kitano2012, Kitano2013}.
Our discussion clarifies what is really the relation between the gluon
and the vector mesons.
%
%
In the UV theory, if we insert a sharp interface of $\theta$ from $0$ to
$2\pi N_f$, there appears an $SU(N_c)_{N_f}$ CS theory whose gauge field
is the gluon field.  If we assume that confinement in the bulk does not
affect the theory on the wall, we find that even at low energy, the
gauge field of the CS theory on the wall is actually the gluon field.
The field $c$ in our description of the wall theory, \cref{wall_v}, is
the gauge field in the level-rank dual of the $SU(N_c)_{N_f}$ CS theory.
In this sense, $c$ can be regarded as the ``dual gluon field.''  
The CS theory of $c$ and the hidden local symmetry of $v$ have the same gauge group,
$U(N_f)$, which makes it possible for them to mix.
Indeed, we find that such a mixing is necessary on the edge of the
pancake as in \cref{wall_v} in order to maintain the gauge
invariance of the external gauge fields, i.e., for consistency.
%

%


Let us go back to the discussion of the phase transition.  When we go to
the finite temperature system, $\eta'$ can cross the cusps of the
potential more easily and frequently at high temperatures.  In this
situation, $c$ can propagate in the four-dimensional space due to
creation of the pancakes, and mixes with $v$ and they are eventually
indistinguishable.
Although a simple linear sigma model is not possible as we discussed
above, the vector mesons can gradually deform into the dual gluons in
this picture.
If this is the correct picture, the pancakes are essential for the
description of the behavior of the vector resonances near the phase
transition.

%

It is interesting to consider the $S^1$ compactified QCD where $\theta$
winds around the $S^1$ direction $N_f$ times.
%
The three-dimensional effective theory for a small radius is known to be
an $SU(N_c)_{N_f}$ CS theory, and for a large radius, the effective
theory is the one we discussed in this paper.
In Ref.~\cite{Kan2019}, it has been claimed that the two limits of the
theory are interpolated by the Higgs mechanism of a $U(N_f)_{-N_c}$ CS
theory with $2N_f$ scalar fields.  Our result is consistent with this.
%
%
At a small radius where the change of $\theta$ is so rapid, $\eta'$
cannot follow the change of $\theta$, which means there appears the
domain wall with $k=N_f$. The $U(N_f)_{-N_c}$ CS theory of $c$ is
realized on the wall, which is naturally identified as the level-rank
dual of the theory in the small radius limit, $SU(N_c)_{N_f}$. The gauge
boson of $SU(N_c)_{N_f}$ is nothing but the gluons.
On the other hand, for a large radius, $\eta'$ follows the change of
$\theta$, and then there is no domain wall. Instead of the CS theory for
$c$, there is a CS Higgs theory for the vector mesons (at least if $m$ in \cref{n} is not equal to 1).
Again, the creation of the pancakes can make the transition smooth by
the mixing between $v$ and $c$ on the edges.
%
%
Although the simple Higgs mechanism does not work in four dimensions,
the transition of the lowest modes, i.e., the three-dimensional
effective theory, can be a smooth transition from the CS Higgs theory of
vector mesons to the CS theory of $c$ as the radius gets smaller.

\section*{Acknowledgements}
We would like to thank Avner Karasik and Shimon Yankielowicz for discussions and useful comments.
The work is supported by JSPS KAKENHI Grant No.~19H00689 (RK and RM), and MEXT KAKENHI Grant No.~18H05542 (RK).

\bibliographystyle{utphys}
\bibliography{../../paper5}

\providecommand{\href}[2]{#2}\begingroup\raggedright\begin{thebibliography}{10}

\bibitem{Hooft1980}
G.~{'t Hooft}, {Naturalness, Chiral Symmetry, and Spontaneous Chiral Symmetry
  Breaking}, \href{http://dx.doi.org/10.1007/978-1-4684-7571-5_9}{NATO Sci.
  Ser. B {\bfseries 59} (1980) 135--157}.

\bibitem{Wess1971}
J.~Wess and B.~Zumino, {Consequences of anomalous ward identities},
  \href{http://dx.doi.org/10.1016/0370-2693(71)90582-X}{Phys. Lett. B
  {\bfseries 37} (1971) 95--97}.

\bibitem{Witten1983a}
E.~Witten, {Global aspects of current algebra},
  \href{http://dx.doi.org/10.1016/0550-3213(83)90063-9}{Nucl. Phys. B
  {\bfseries 223} (1983) 422--432}.

\bibitem{Hooft1974}
G.~{'t Hooft}, {A planar diagram theory for strong interactions},
  \href{http://dx.doi.org/10.1016/0550-3213(74)90154-0}{Nucl. Phys. B
  {\bfseries 72} (1974) 461--473}.

\bibitem{Witten1979}
E.~Witten, {Current algebra theorems for the U(1) “Goldstone boson”},
  \href{http://dx.doi.org/10.1016/0550-3213(79)90031-2}{Nucl. Phys. B
  {\bfseries 156} (1979) 269--283}.

\bibitem{Veneziano1979}
G.~Veneziano, {U(1) without instantons},
  \href{http://dx.doi.org/10.1016/0550-3213(79)90332-8}{Nucl. Phys. B
  {\bfseries 159} (1979) 213--224}.

\bibitem{Witten1979a}
E.~Witten, {Baryons in the 1/N expansion},
  \href{http://dx.doi.org/10.1016/0550-3213(79)90232-3}{Nucl. Phys. B
  {\bfseries 160} (1979) 57--115}.

\bibitem{Witten1983}
E.~Witten, {Current algebra, baryons, and quark confinement},
  \href{http://dx.doi.org/10.1016/0550-3213(83)90064-0}{Nucl. Phys. B
  {\bfseries 223} (1983) 433--444}.

\bibitem{Skyrme1961}
T.~Skyrme, {A non-linear field theory},
  \href{http://dx.doi.org/10.1098/rspa.1961.0018}{Proc. R. Soc. London. Ser. A.
  Math. Phys. Sci. {\bfseries 260} (1961) 127--138}.

\bibitem{Kapustin2014b}
A.~Kapustin and R.~Thorngren, {Anomalous Discrete Symmetries in Three
  Dimensions and Group Cohomology},
  \href{http://dx.doi.org/10.1103/PhysRevLett.112.231602}{Phys. Rev. Lett.
  {\bfseries 112} (2014) 231602},
  \href{http://arxiv.org/abs/1403.0617}{{\ttfamily arXiv:1403.0617}}.

\bibitem{Kapustin2014}
A.~Kapustin and N.~Seiberg, {Coupling a QFT to a TQFT and duality},
  \href{http://dx.doi.org/10.1007/JHEP04(2014)001}{JHEP {\bfseries 04} (2014)
  1}, \href{http://arxiv.org/abs/1401.0740}{{\ttfamily arXiv:1401.0740}}.

\bibitem{Gaiotto2015}
D.~Gaiotto, A.~Kapustin, N.~Seiberg, and B.~Willett, {Generalized global
  symmetries}, \href{http://dx.doi.org/10.1007/JHEP02(2015)172}{JHEP {\bfseries
  02} (2015) 1--62}, \href{http://arxiv.org/abs/1412.5148}{{\ttfamily
  arXiv:1412.5148}}.

\bibitem{Gaiotto2017}
D.~Gaiotto, A.~Kapustin, Z.~Komargodski, and N.~Seiberg, {Theta, time reversal
  and temperature}, \href{http://dx.doi.org/10.1007/JHEP05(2017)091}{JHEP
  {\bfseries 2017} (2017) 91}.

\bibitem{Hsin2019}
P.-S. Hsin, H.~T. Lam, and N.~Seiberg, {Comments on one-form global symmetries
  and their gauging in 3d and 4d},
  \href{http://dx.doi.org/10.21468/SciPostPhys.6.3.039}{SciPost Phys.
  {\bfseries 6} (2019) 039}, \href{http://arxiv.org/abs/1812.04716}{{\ttfamily
  arXiv:1812.04716}}.

\bibitem{Kitano2017}
R.~Kitano, T.~Suyama, and N.~Yamada, {$\theta=\pi$ in $SU(N)/\mathbb{Z}_N$
  gauge theories}, \href{http://dx.doi.org/10.1007/JHEP09(2017)137}{JHEP
  {\bfseries 09} (2017) 137}, \href{http://arxiv.org/abs/1709.04225}{{\ttfamily
  arXiv:1709.04225}}.

\bibitem{Cordova2020}
C.~Cordova, D.~Freed, H.~T. Lam, and N.~Seiberg, {Anomalies in the space of
  coupling constants and their dynamical applications I},
  \href{http://dx.doi.org/10.21468/SciPostPhys.8.1.001}{SciPost Phys.
  {\bfseries 8} (2020) 001}, \href{http://arxiv.org/abs/1905.09315}{{\ttfamily
  arXiv:1905.09315}}.

\bibitem{Cordova2020a}
C.~Cordova, D.~Freed, H.~T. Lam, and N.~Seiberg, {Anomalies in the space of
  coupling constants and their dynamical applications II},
  \href{http://dx.doi.org/10.21468/SciPostPhys.8.1.002}{SciPost Phys.
  {\bfseries 8} (2020) 002}, \href{http://arxiv.org/abs/1905.13361}{{\ttfamily
  arXiv:1905.13361}}.

\bibitem{Komargodski2018a}
Z.~Komargodski, {Baryons as Quantum Hall Droplets},
  \href{http://arxiv.org/abs/1812.09253}{{\ttfamily arXiv:1812.09253}}.

\bibitem{Ma2019}
Y.~L. Ma, M.~A. Nowak, M.~Rho, and I.~Zahed, {Baryon as a Quantum Hall Droplet
  and the Quark-Hadron Duality},
  \href{http://dx.doi.org/10.1103/PhysRevLett.123.172301}{Phys. Rev. Lett.
  {\bfseries 123} (2019) 172301}.

\bibitem{Karasik2020}
A.~Karasik, {Skyrmions, Quantum Hall Droplets, and one current to rule them
  all}, \href{http://arxiv.org/abs/2003.07893}{{\ttfamily arXiv:2003.07893}}.

\bibitem{Karasik2020a}
A.~Karasik, {Vector dominance, one flavored baryons, and QCD domain walls from
  the "hidden" Wess-Zumino term},
  \href{http://arxiv.org/abs/2010.10544}{{\ttfamily arXiv:2010.10544}}.

\bibitem{Kan2019}
N.~Kan, R.~Kitano, S.~Yankielowicz, and R.~Yokokura, {From 3d dualities to
  hadron physics}, \href{http://arxiv.org/abs/1909.04082}{{\ttfamily
  arXiv:1909.04082}}.

\bibitem{Ma2020}
Y.-L. Ma and M.~Rho, {Dichotomy of Baryons as Quantum Hall Droplets and
  Skyrmions In Compact-Star Matter},
  \href{http://arxiv.org/abs/2009.09219}{{\ttfamily arXiv:2009.09219}}.

\bibitem{Seiberg1995}
N.~Seiberg, {Electric-magnetic duality in supersymmetric non-Abelian gauge
  theories}, \href{http://dx.doi.org/10.1016/0550-3213(94)00023-8}{Nucl. Phys.
  B {\bfseries 435} (1995) 129--146},
  \href{http://arxiv.org/abs/hep-th/9411149}{{\ttfamily hep-th/9411149}}.

\bibitem{Seiberg2001}
N.~Seiberg, {The Power of Duality -- Exact Results in 4D SUSY Field Theory},
  \href{http://dx.doi.org/10.1142/S0217751X01005705}{Int. J. Mod. Phys. A
  {\bfseries 16} (2001) 4365--4376},
  \href{http://arxiv.org/abs/hep-th/9506077}{{\ttfamily hep-th/9506077}}.

\bibitem{Harada1999}
M.~Harada and K.~Yamawaki, {Conformal Phase Transition and Fate of the Hidden
  Local Symmetry in Large $N_f$ QCD},
  \href{http://dx.doi.org/10.1103/PhysRevLett.83.3374}{Phys. Rev. Lett.
  {\bfseries 83} (1999) 3374--3377},
  \href{http://arxiv.org/abs/hep-ph/9906445}{{\ttfamily hep-ph/9906445}}.

\bibitem{Komargodski2011}
Z.~Komargodski, {Vector mesons and an interpretation of Seiberg duality},
  \href{http://dx.doi.org/10.1007/JHEP02(2011)019}{JHEP {\bfseries 02} (2011)
  1--21}.

\bibitem{Kitano2011}
R.~Kitano, {Hidden local symmetry and color confinement},
  \href{http://dx.doi.org/10.1007/JHEP11(2011)124}{JHEP {\bfseries 11} (2011)
  124}, \href{http://arxiv.org/abs/1109.6158}{{\ttfamily arXiv:1109.6158}}.

\bibitem{Abel2012}
S.~Abel and J.~Barnard, {Seiberg duality versus hidden local symmetry},
  \href{http://dx.doi.org/10.1007/JHEP05(2012)044}{JHEP {\bfseries 05} (2012)
  044}, \href{http://arxiv.org/abs/1202.2863}{{\ttfamily arXiv:1202.2863}}.

\bibitem{Kitano2012}
R.~Kitano, M.~Nakamura, and N.~Yokoi, {Making confining strings out of mesons},
  \href{http://dx.doi.org/10.1103/PhysRevD.86.014510}{Phys. Rev. D {\bfseries
  86} (2012) 014510}, \href{http://arxiv.org/abs/1202.3260}{{\ttfamily
  arXiv:1202.3260}}.

\bibitem{Kitano2013}
R.~Kitano and N.~Yokoi, {Quark confinement via magnetic color-flavor locking},
  \href{http://dx.doi.org/10.1007/JHEP11(2013)129}{JHEP {\bfseries 11} (2013)
  129}, \href{http://arxiv.org/abs/1308.0093}{{\ttfamily arXiv:1308.0093}}.

\bibitem{Bando1985}
M.~Bando, T.~Kugo, S.~Uehara, K.~Yamawaki, and T.~Yanagida, {Is the $\rho$
  meson a dynamical gauge boson of hidden local symmetry?},
  \href{http://dx.doi.org/10.1103/PhysRevLett.54.1215}{Phys. Rev. Lett.
  {\bfseries 54} (1985) 1215--1218}.

\bibitem{Fujiwara1985}
T.~Fujiwara, T.~Kugo, H.~Terao, S.~Uehara, and K.~Yamawaki, {Non-Abelian
  Anomaly and Vector Mesons as Dynamical Gauge Bosons of Hidden Local
  Symmetries}, \href{http://dx.doi.org/10.1143/ptp.73.926}{Prog. Theor. Phys.
  {\bfseries 73} (1985) 926--941}.

\bibitem{Hidaka2020a}
Y.~Hidaka, M.~Nitta, and R.~Yokokura, {Global 3-group symmetry and 't Hooft
  anomalies in axion electrodynamics},
  \href{http://arxiv.org/abs/2009.14368}{{\ttfamily arXiv:2009.14368}}.

\bibitem{Hidaka2020b}
Y.~Hidaka, M.~Nitta, and R.~Yokokura, {Higher-form symmetries and 3-group in
  axion electrodynamics},
  \href{http://dx.doi.org/10.1016/j.physletb.2020.135672}{Phys. Lett. B
  {\bfseries 808} (2020) 135672}.

\bibitem{Witten1979c}
E.~Witten, {Dyons of charge e$\theta$/2$\pi$},
  \href{http://dx.doi.org/10.1016/0370-2693(79)90838-4}{Phys. Lett. B
  {\bfseries 86} (1979) 283--287}.

\bibitem{Corrigan1976}
E.~Corrigan, D.~Olive, D.~Fairlie, and J.~Nuyts, {Magnetic monopoles in SU(3)
  gauge theories}, \href{http://dx.doi.org/10.1016/0550-3213(76)90391-6}{Nucl.
  Phys. B {\bfseries 106} (1976) 475--492}.

\bibitem{Gaiotto2018}
D.~Gaiotto, Z.~Komargodski, and N.~Seiberg, {Time-Reversal Breaking in QCD$_4$,
  Walls, and Dualities in 2+1 Dimensions},
  \href{http://dx.doi.org/10.1007/JHEP01(2018)110}{JHEP {\bfseries 01} (2018)
  110}, \href{http://arxiv.org/abs/1708.06806}{{\ttfamily arXiv:1708.06806}}.

\bibitem{Tanizaki2018}
Y.~Tanizaki, {Anomaly constraint on massless QCD and the role of Skyrmions in
  chiral symmetry breaking},
  \href{http://dx.doi.org/10.1007/JHEP08(2018)171}{JHEP {\bfseries 08} (2018)
  171}, \href{http://arxiv.org/abs/1807.07666}{{\ttfamily arXiv:1807.07666}}.

\bibitem{Anber2019}
M.~M. Anber and E.~Poppitz, {On the baryon-color-flavor (BCF) anomaly in
  vector-like theories}, \href{http://dx.doi.org/10.1007/JHEP11(2019)063}{JHEP
  {\bfseries 11} (2019) 1--13},
  \href{http://arxiv.org/abs/1909.09027}{{\ttfamily arXiv:1909.09027}}.

\bibitem{Dierigl2014}
M.~Dierigl and A.~Pritzel, {Topological model for domain walls in
  (super-)Yang-Mills theories},
  \href{http://dx.doi.org/10.1103/PhysRevD.90.105008}{Phys. Rev. D {\bfseries
  90} (2014) 105008}, \href{http://arxiv.org/abs/1405.4291}{{\ttfamily
  arXiv:1405.4291}}.

\bibitem{Moore1989}
G.~Moore and N.~Seiberg, {Taming the conformal zoo},
  \href{http://dx.doi.org/10.1016/0370-2693(89)90897-6}{Phys. Lett. B
  {\bfseries 220} (1989) 422--430}.

\bibitem{Elitzur1989}
S.~Elitzur, G.~Moore, A.~Schwimmer, and N.~Seiberg, {Remarks on the canonical
  quantization of the Chern-Simons-Witten theory},
  \href{http://dx.doi.org/10.1016/0550-3213(89)90436-7}{Nucl. Phys. B
  {\bfseries 326} (1989) 108--134}.

\bibitem{Furui1986}
S.~Furui, R.~Kobayashi, and K.~Ujiie, {A Comment on the $K_{l4}$ Decay in the
  Gauged Nonlinear Sigma Model with the Wess-Zumino Condition},
  \href{http://dx.doi.org/10.1143/PTP.76.963}{Prog. Theor. Phys. {\bfseries 76}
  (1986) 963--965}.

\end{thebibliography}\endgroup
\end{document}